\documentclass[12pt,epsf]{article}
\usepackage[pctex32]{graphics}
\usepackage{cancel}
\makeatletter

\@addtoreset{equation}{section}
\newcommand{\be}{\begin{eqnarray}}
\newcommand{\ee}{\end{eqnarray}}
\newcommand{\ba}{\begin{array}}
\newcommand{\ea}{\end{array}}
\newcommand{\nn}{\nonumber}

\makeatletter \@addtoreset{equation}{section} \makeatother

\begin{document}
\vspace{1cm}
\begin{center}
~\\~\\~\\
{\bf  \LARGE General Lagrangian of Non-Covariant Self-dual Gauge Field}
\vspace{1cm}

                      Wung-Hong Huang\\
                       Department of Physics\\
                       National Cheng Kung University\\
                       Tainan, Taiwan\\

\end{center}
\vspace{1cm}
\begin{center}{\bf  \Large ABSTRACT } \end{center}
We present the general formulation of  non-covariant  Lagrangian of self-dual gauge theory.  After specifying the parameters therein the previous Lagrangian in the decomposition of spacetime into $6=D_1+D_2$ and $6=D_1+D_2+D_3$ can be obtained.  The self-dual property of the general Lagrangian is proved in detail.  We furthermore  show that the new non-covariant actions give field equations with 6d Lorentz invariance.  The method can be straightforward extended to any dimension and we also give a short discussion about the 10D self-dual gauge theory.
 
\vspace{4cm}
\begin{flushleft}
*E-mail:  whhwung@mail.ncku.edu.tw
\end{flushleft}
\newpage
\tableofcontents
\section{Introduction}
The problem in Lagrangian description of chiral p-forms, i.e. antisymmetric boson fields with self-dual had been known before thirty years ago. As first noted by Marcus and Schwarz [1], the manifest duality and spacetime covariance do not like to live in harmony with each other in one action. 

   Historically, the non-manifestly spacetime covariant action for self-dual  0-form was proposed by Floreanini and Jackiw [2], which is then generalized to p-form by Henneaux and Teitelboim [3].  In general the field strength of chiral p-form $A_{1\cdot\cdot\cdot p}$ is split into electric density $ {\cal E}^{i_1\cdot\cdot\cdot i_{p+1}}$ and magnetic density $ {\cal B}^{i_1\cdot\cdot\cdot i_{p+1}}$:
\be  {\cal E}_{i_1\cdot\cdot\cdot i_{p+1}} &\equiv& F_{i_1\cdot\cdot\cdot i_{p+1}}\equiv \partial_{[{i_1}}A_{i_2\cdot\cdot\cdot i_{p+1}]}\\
 {\cal B}^{i_1\cdot\cdot\cdot i_{p+1}} &\equiv &{1\over (p+1)!}\epsilon^{i_1\cdot\cdot\cdot i_{2p+2}}  F_{i_{p+2}\cdot\cdot\cdot i_{2p+2}}\equiv \tilde F^{i_1\cdot\cdot\cdot i_{p+1}} 
\ee
in which $\tilde F$ is the dual form of $F$.  The Lagrangian is described by
\be  L= -{1\over p!} {\vec{\cal B}}\cdot ({\vec{\cal E}}-{\vec{\cal B}})={1\over p!}  \tilde F_{i_1\cdot\cdot\cdot i_{p+1}} {\cal F}^{i_1\cdot\cdot\cdot i_{p+1}}
\ee
in which we define
\be {\cal F}^{i_1\cdot\cdot\cdot i_{p+1}}
\equiv \tilde F^{i_1\cdot\cdot\cdot i_{p+1}}- F^{i_1\cdot\cdot\cdot i_{p+1}}
\ee
Note that in order for self-dual fields to exist, i.e. $\tilde F=F$, the field strength $F$ and dual field  strength $\tilde F$ should have the same number of component. As the double dual on field strength shall give the original field strength the spacetime dimension have to be 2 modulo 4. 

Four years ago,  a new non-covariant Lagrangian formulation of a chiral 2-form gauge field in 6D, called as $6=3+3$ decomposition,  was derived in [4] from the Bagger-Lambert-Gustavsson (BLG) model [5].   Later, a general non-covariant Lagrangian formulation of  self-dual gauge theories in diverse dimensions was constructed [6].  In this general formulation the $6=2+4$ decomposition of  Lagrangian was found.  

In the last year we have constructed a new kind of non-covariant actions of self-dual 2-form gauge theory in the decomposition of  $6=D_1+D_2+D_3$ [7].  In this paper we will present the general formulation of  non-covariant  Lagrangian of self-dual gauge theory.

   In section II we first present the general non-covariant  Lagrangian of self-dual gauge theory.  We see that after specifying the parameters therein the all known Lagrangian in the decomposition of  $6=D_1+D_2$ [6] and $6=D_1+D_2+D_3$ [7] can be obtained.  In section III  We discuss some properties which are crucial to formulate the general  Lagrangian of  non-covariant forms of self-dual gauge theory.  We then prove in detail the self-dual property of the general Lagrangian.  In section IV we follow Perry and Schwarz [8] to show that the general non-covariant Lagrangian gives field equations with 6d Lorentz invariance.   Our prescription can be straightforward extended to any dimension and we also give a brief description about the 10D self-dual gauge theory in section V.  Last section is devoted to a short conclusion.
\section {Lagrangian of Self-dual Gauge Fields in Simple Decomposition}
In this section we first present the general non-covariant  Lagrangian $L_G$ of self-dual gauge theory in (2.2) and table 1. Then we collect all know non-covariant  Lagrangian of self-dual gauge theory in six dimension [6,7] and compare them with $L_G$.   Table 2 and table 3 are just those in our previous paper [7], while for convenience we reproduce them in this paper.  We will see that, after specifying the parameters in $L_G$  the previous Lagrangian in the decomposition of spacetime into $6=D_1+D_2$ and $6=D_1+D_2+D_3$ can be obtained. \\

 To begin with, let us first define a  function $L_{ijk}$ :
\be L_{ijk}&\equiv& \tilde F_{ijk}\times(F^{ijk}-\tilde F^{ijk}), ~~without~summation~over~indices~i,j,k
\ee
which is useful in the following formulations.

   In terms of $L_{ijk}$ the most general non-covariant  Lagrangian of self-dual gauge theory  we found is  
\be L_G(\alpha_i) =  \sum\limits_a L_{12 a}+
({1\over2}+{\alpha_1\over2}) L_{134}+ ({1\over2}-{\alpha_1\over2}) L_{256}+({1\over2}+{\alpha_2\over2}) L_{135}+ ({1\over2}-{\alpha_2\over2}) L_{246}\nn\\
({1\over2}+{\alpha_3\over2}) L_{136}+ ({1\over2}-{\alpha_3\over2}) L_{245}+ ({1\over2}+{\alpha_4\over2}) L_{145}+ ({1\over2}-{\alpha_4\over2}) L_{236}\nn\\
({1\over2}+{\alpha_5\over2}) L_{146}+ ({1\over2}-{\alpha_5\over2}) L_{235}+ ({1\over2}+{\alpha_6\over2}) L_{156}+ ({1\over2}-{\alpha_6\over2}) L_{234}\nn \\
\ee
Let us make three comments about above Lagrangian.

First, From table 1 we see that $L_{G}$ does not picks up $L_{456}$, $L_{356}$, $ L_{346}$ nor $L_{345}$, which is denoted as ${\cancel L_{abc}}$.  This can ensure to the existence of  gauge symmetry $\delta A_{12}= \Phi_{12}$, which is crucial to prove the self-duality of $L_G$, as shown in next section. 

Next, We have chosen the coefficient before $L_{12 a}$ to be one.  This is because that the overall constant of $L_G$ does not affect the self-duality.

Third, we choose coefficient $({1\over2}+{\alpha_1\over2})$ before $L_{134}$ while choose coefficient $({1\over2}-{\alpha_1\over2})$ before $L_{256}$. This can ensure that adding the two coefficient $({1\over2}+{\alpha_1\over2})$ + $({1\over2}-{\alpha_1\over2})=1$, which give a proper normalization.  This property is also crucial to prove the self-duality of $L_G$, as shown in next section. 
\\\\\\
{Table 1: Lagrangian in the general decompositions: $D=6$.}\\\\
\scalebox{1}{\hspace{3cm}\includegraphics{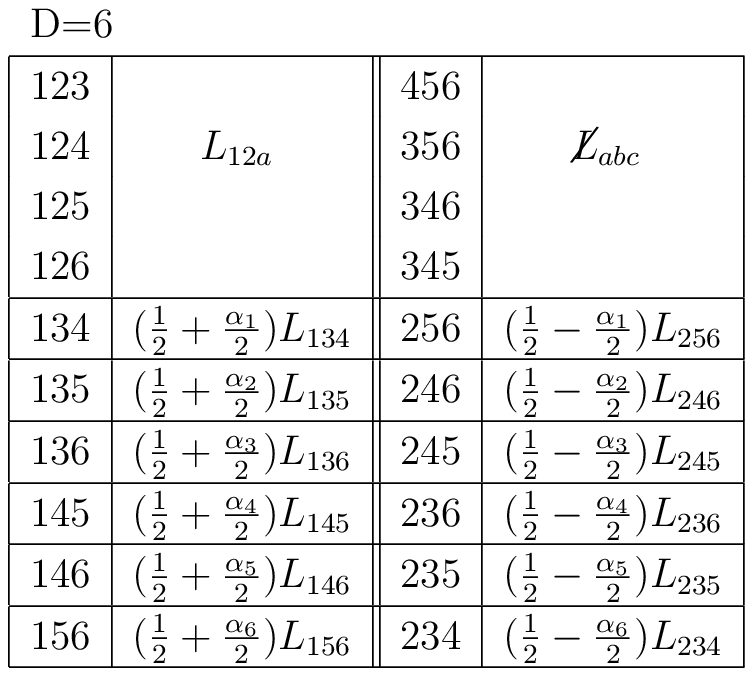}}
\subsection {Lagrangian in Decomposition:  $6=1+5$}
In the (1+5) decomposition the spacetime index $A= (1,\cdot\cdot\cdot,6)$ is decomposed as $A=(1,\dot a)$, with $\dot a=(2,\cdot\cdot\cdot,6)$. Then $L_{ABC}=(L_{1\dot a\dot b}, L_{\dot a\dot b\dot c})$.   In terms of $L_{ABC}$, the Lagrangian is expressed as [6]
\be L_{1+5} = -{1\over4} \sum L_{1\dot a\dot b}=- {1\over4} \tilde F_{1\dot a\dot b}(F^{1\dot a\dot b} -\tilde F^{1\dot a\dot b}),~~has~summation~over~\dot a~\dot b
\ee
From table 2 we see that $L_{1+5}$ picks up only $ L_{1\dot a\dot b}$. Note that that $L_G(\alpha_i=1)=L_{1+5}$. 
\\

{Table 2: Lagrangian in various decompositions: $D=D_1+D_2$.}\\\\
\scalebox{1}{\hspace{0.5cm}\includegraphics{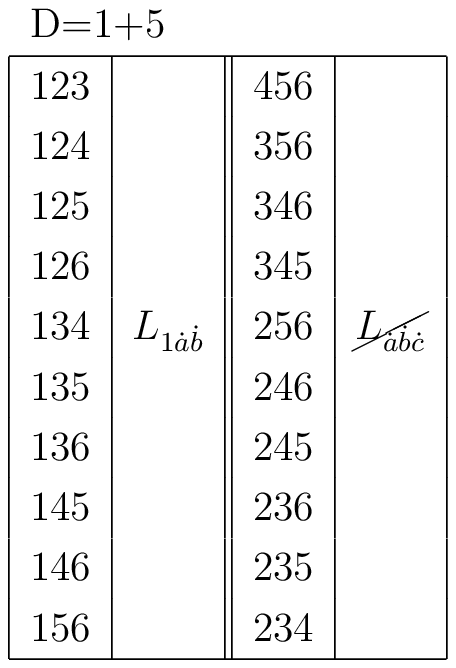}}\scalebox{1}{\hspace{0.5cm}\includegraphics{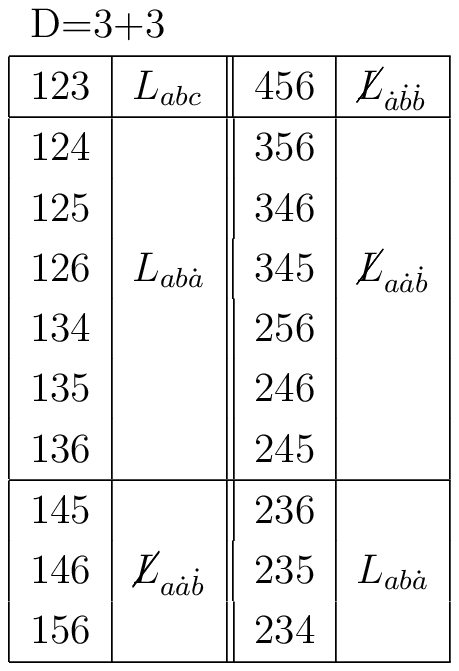}}\scalebox{1}{\hspace{0.5cm}\includegraphics{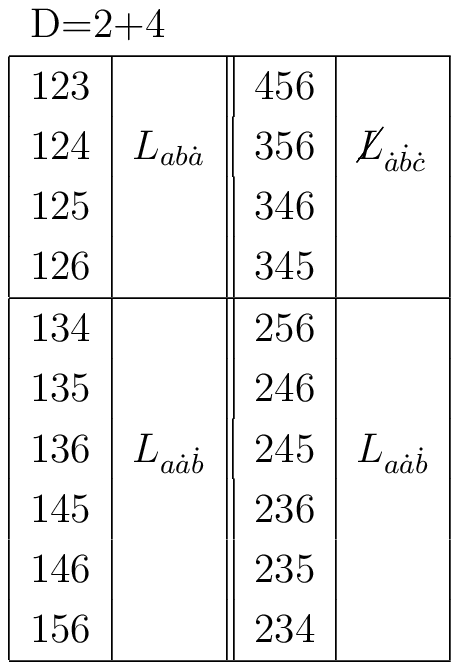}}
\\
\subsection {Lagrangian in Decomposition:  $6=2+4$}
In the (2+4) decomposition [6] the spacetime index $A$ is decomposed as $A=(a,\dot a)$, with $a=(1,2)$ and $\dot a=(3,\cdot\cdot\cdot,6)$. Then $L_{ABC}=(L_{ab\dot a}, L_{a\dot a\dot b},L_{\dot a\dot b\dot c})$.  From table 2 it is easy to see that in terms of $L_{ABC}$ the Lagrangian can be expressed as [7]
\be L_{2+4} = -{1\over4} \Big( \sum L_{ab\dot a}+{1\over2}  \sum L_{a\dot a\dot b}\Big)
\ee
The ${1\over2}$ factor before $L_{a\dot a\dot b}$ arising from the  property that $L_{a\dot a\dot b}$ contains both of left-hand side element and right-hand side  element (for example, it includes $L_{134}$ and $L_{256}$), thus there is double counting. Note that that $L_G(\alpha_i=0)=L_{2+4}$. 
\subsection {Lagrangian in Decomposition:  $6=3+3$}
In the (3+3) decomposition [6] the spacetime index $A$ is decomposed as $A=(a,\dot a)$, with $a=(1,2,3)$ and $\dot a=(4,5,6)$. Then $L_{ABC}=(L_{abc}, L_{ab\dot a}, L_{a\dot a\dot b}, L_{\dot a\dot b\dot c})$.  Using table 2 it is easy to see that in terms of $L_{ABC}$ the Lagrangian can be expressed as 
\be L_{3+3} = -{1\over12} \Big( \sum L_{abc}+3  \sum L_{ab\dot a}\Big)
\ee
The ``3'' factor before $L_{ab\dot a}$ arising from the property that we have to include three kinds of $L_{ijk}$ : $L_{ab\dot a}$, $L_{a\dot a b}$ and $L_{\dot a ab}$.  Note that $L_G(\alpha_1=\alpha_2=\alpha_3=1;\alpha_4=\alpha_5=\alpha_6=-1)=L_{3+3}$. 

Self-dual property of above decomposition  had been proved in [6].
\subsection {Lagrangian in Decomposition:  $6=1+1+4$}
In the (1+1+4) decomposition the spacetime index $A$ is decomposed as $A=(1,2,\dot a)$, with $\dot a=(3,4,5,6)$, and  $L_{ABC}=(L_{12\dot a}, L_{\dot a \dot b\dot c}, L_{1\dot a\dot b}, L_{2\dot a\dot b})$.  From table 3 it is easy to see that, in terms of $L_{ABC}$, the Lagrangian can be expressed as [7]
\be L_{1+1+4} = 6 \sum L_{12\dot a}+ {3(1+\alpha)\over2}\sum L_{1\dot a\dot b}+{3(1-\alpha)\over2}\sum L_{2\dot a\dot b}
\ee
We neglect overall constant in Lagrangian, which is irrelevant to the self-duality.

It is easy to see that $L_{1+1+4}$ in the case of $\alpha =0$ is just $L_{2+4}$,  in the case of $\alpha =1$ is just $L_{1+5}$, and in the case of $\alpha =-1$  is just $L_{1+5}$ while  exchanging indices  1 and 2, as can be seen from table 2. Note that $L_G(\alpha_i=\alpha)=L_{1+1+4}$.  
\\\\\\\\
{Table 3: Lagrangian in various decompositions: $D=D_1+D_2+D3$.}\\\\
\scalebox{1}{\hspace{0.5cm}\includegraphics{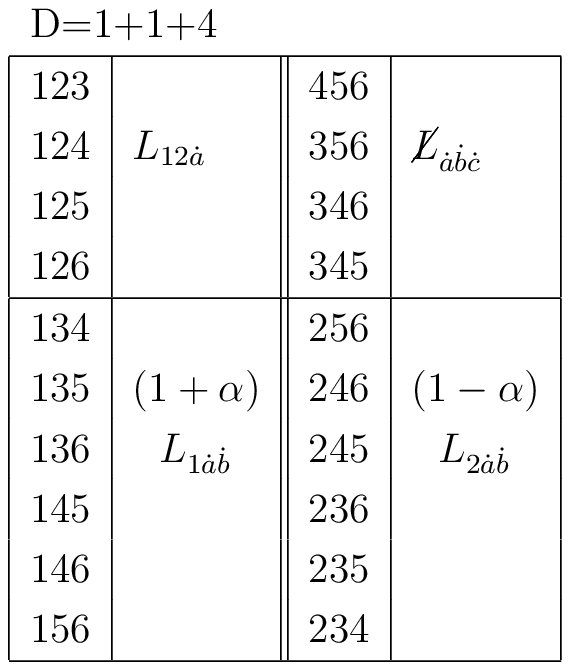}}\scalebox{1}{\hspace{0.5cm}\includegraphics{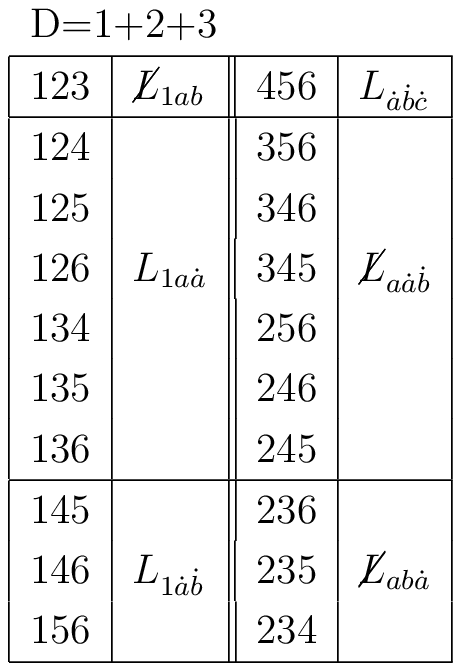}}\scalebox{1}{\hspace{0.5cm}\includegraphics{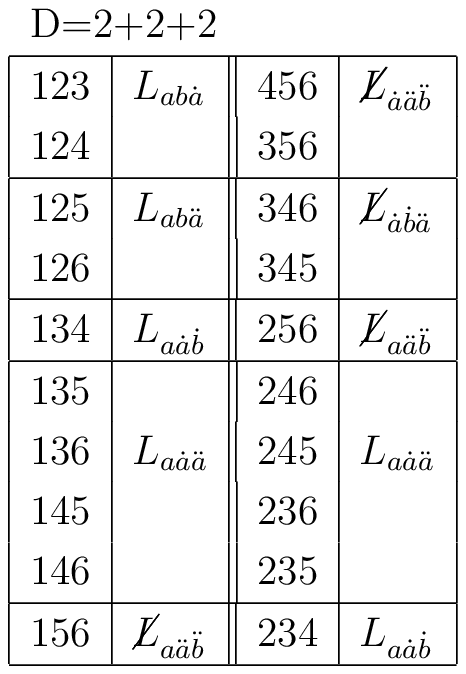}}
\\
\subsection {Lagrangian in Decomposition:  $6=1+2+3$}
In the (1+2+3) decomposition the spacetime index $A$ is decomposed as $A=(1,a ,\dot a)$, with $a=(2,3)$, $\dot a=(4,5,6)$, and  $L_{ABC}=(L_{1ab}$, $L_{1a\dot a}$, $L_{1\dot a\dot b}$, $L_{\dot a\dot b\dot c}$, $L_{a\dot a\dot b}$, $L_{ab\dot a})$.  From table 3 it is easy to see that, in terms of $L_{ABC}$, the Lagrangian can be expressed as [7]
\be L_{1+2+3} = \sum L_{\dot a\dot b\dot c}+ 6 \sum L_{1a\dot a}+3 \sum L_{1\dot a\dot b}
\ee
Choosing $L_{1ab}+L_{1a\dot a}+L_{1\dot a\dot b}$ is just $L_{1+5}$, and choosing $L_{1ab}+L_{1 a\dot a}+L_{ab\dot a}$ is just $L_{3+3}$, as can be seen from table 2.  Note that, exchanging indices  2 with 5 and 3 with 6 then  $L_G(\alpha_1=\alpha_2=\alpha_3=\alpha_4=\alpha_5=1,\alpha_6=-1)=L_{1+2+3}$.
\subsection {Lagrangian in Decomposition:  $6=2+2+2$}
In the (2+2+2) decomposition the spacetime index $A$ is decomposed as $A=(a,\dot a,\ddot a)$, with $a=(1,2)$, $\dot a=(3,4)$ and $\ddot a=(5,6)$.  Now, from table 3 we see that  $L_{ABC}=(L_{a\dot a\dot b}$, $L_{\dot a\ddot a\ddot b}$, $L_{ab\ddot a}$, $L_{\dot a\dot b\ddot a}$, $L_{a\dot a\dot b}$, $L_{a\ddot a\ddot b}$, $L_{a\dot a\ddot a})$. Then, in terms of $L_{ABC}$ the Lagrangian can be expressed as [7]
\be L_{2+2+2} = \sum L_{ab\dot a}+  \sum L_{ab\ddot a}+  \sum L_{a\dot a\dot b}+\sum L_{a\dot a\ddot a}
\ee
Note that $L_G(\alpha_1=1, \alpha_2=\alpha_3=\alpha_4=\alpha_5=0,\alpha_6=-1)=L_{2+2+2}$.

Self-dual property of above decompositions  had been proved in [6, 7].
\section {General Lagrangian of Self-dual Gauge Fields}
In order to understand how to find the general formulation we need to find some constrains in formulating the non-covariant  Lagrangian of self-dual gauge theory [6,7].  

\subsection {General Lagrangian and Gauge Symmetry}
   From previous studies [6,7] we see that the proof of the self-dual property has used a {\bf Gauge symmetry}.  The property tells us that terms involved $A_{12}$, for example, only through total derivative terms and we have gauge symmetry
\be   \delta A_{12}&=&\Phi_{12}
\ee
for arbitrary functions $\Phi_{12}$.  In order to have this property we must not choose the dual transformationof $L_{12a}$ in table 1, where a=(3,4,5,6).  Thus the general non-covariant  Lagrangian of self-dual gauge theory is  
\be L =  \sum\limits_a L_{12 a}+
C_1({1\over2}+{\alpha_1\over2}) L_{134}+ C_1({1\over2}-{\tilde \alpha_1\over2}) L_{256}+C_2({1\over2}+{\alpha_2\over2}) L_{135}+ C_2({1\over2}-{\tilde \alpha_2\over2}) L_{246}\nn\\
C_3({1\over2}+{\alpha_3\over2}) L_{136}+ C_3({1\over2}-{\tilde \alpha_3\over2}) L_{245}+ C_4({1\over2}+{\alpha_4\over2}) L_{145}+ C_4({1\over2}-{\tilde \alpha_4\over2}) L_{236}\nn\\
C_5({1\over2}+{\alpha_5\over2}) L_{146}+ C_5({1\over2}-{\tilde \alpha_5\over2}) L_{235}+ C_6({1\over2}+{\alpha_6\over2}) L_{156}+ C_6({1\over2}-{\tilde \alpha_6\over2}) L_{234}\nn\\
\ee
in which we have let the coefficient before $L_{12 a}$ to be one as  the overall constant of $L$ does not affect the self-duality.  We will now show that above  Lagrangian has desired gauge symmetry, and after proper choosing the parameters $C_i$  it becomes $L_G$ in (2.2) and the associated field strength has self-dualtiy property.\\

First, the variation of the action gives 
\be {\delta L\over \delta A_{12}}&=&-\partial_{3} \tilde F^{312}-\partial_{4} \tilde F^{412}-\partial_{5} \tilde F^{512}-\partial_{6} \tilde F^{612}=\partial_{ a}\tilde F^{a12}=0 
\ee
which is identically zero.  This means that terms involved $A_{12}$ only through total derivative terms and we have gauge symmetry
\be   \delta A_{12}&=&\Phi_{12}
\ee
for arbitrary functions $\Phi_{12}$.
\subsection {Self-duality}
Next, simply using above gauge symmetry does not guarantee that the Lagrangian has self-dual property.  Let us find the another constrain.

The variation of the Lagrangian $L$ gives 
\be  0={\delta L\over \delta A_{34}}&=&-\Big[C_1\partial_1\tilde F^{134}+C_6\partial_2\tilde F^{234}+\partial_5\tilde F^{534}+\partial_6\tilde F^{634}\Big]\nn\\
&&+C_1(1-{{\alpha_1+\tilde \alpha_1}\over2})\partial_1\tilde F^{134} -C_1(1-\tilde \alpha_1)\partial_1F^{134}\nn\\
&&+C_6(1-{{\alpha_6+\tilde \alpha_6}\over2})\partial_2\tilde F^{234} -C_6(1-\tilde \alpha_6)\partial_2F^{234}\nn\\
&&+2(\partial_5{\cal F}^{534}+\partial_6{\cal F}^{634})\nn\\
\ee
Now, if we require each $L_{i,j,k}$ has a same normalization then $C_i=1$. Under this constrain we find that
\be
 0={\delta L\over \delta A_{34}}&=&(1-{{\alpha_1+\tilde \alpha_1}\over2})\partial_1\tilde F^{134} -(1-\tilde \alpha_1)\partial_1F^{134}\nn\\
&&+(1-{{\alpha_6+\tilde \alpha_6}\over2})\partial_2\tilde F^{234} -(1-\tilde \alpha_6)\partial_2F^{234}\nn\\
&&+2(\partial_5{\cal F}^{534}+\partial_6{\cal F}^{634})
\ee
where we have used the property $\partial_a\tilde F^{a34}=0$. 

To procced, we can from table 2 and table 3 see that, for example, the difference between the Lagrangian in decomposition $D=2+4$ and $D=1+5$ is that we have chosen left-hand (electric) part and right-hand (magnetic) part in $D=2+4$, while in $D=1+5$ we choose only left-hand (electric) part.  However, in the self-dual theory the electric part is equal to magnetic part.   Thus the Lagrangian choosing electric part is equivalent to that choosing magnetic part.  In the decomposition into different direct-product of  spacetime one can  choose different part of $L_{ijk}$ to mixing to each other and we have many kind of decomposition.  This observation lead us to find more decomposition $6=D_1+D_2+D_3$ in [7].   

This property can be applied to find the more general formulation of non-covariant  Lagrangian of self-dual gauge theory. Thus the another constrain is that 
\be
\Big({1\over2}+{\alpha_i\over2}\Big)+\Big({1\over2}-{\tilde \alpha_i\over2}\Big)=1~~~\Rightarrow~~\alpha_i=\tilde\alpha_i
\ee
From now on we will use this property and Lagrangian $L$ becomes $L_G$ in (2.2). 

 Thus
\be
 0={\delta L\over \delta A_{34}}&=&(1-\alpha_1)\partial_1 {\cal F}^{134}+(1+\alpha_6)\partial_2 {\cal F}^{234}+2\partial_5 {\cal F}^{534}+2\partial_6 {\cal F}^{634}\nn\\
&=&\partial_1 \bar {\cal F}^{134}+\partial_2 \bar {\cal F}^{234}+\partial_5 \bar {\cal F}^{534}+\partial_6 \bar {\cal F}^{634}\ee
in which we have normalized each $\cal F$ by the associated factor $(1-\alpha_i)$ or 2 for convenience. 

Similary, we have the relations
\be
0={\delta L\over \delta A_{35}}&=&\partial_1 \bar {\cal F}^{135}+\partial_2 \bar {\cal F}^{235}+\partial_4 \bar {\cal F}^{435}+\partial_6 \bar {\cal F}^{635}\\
0={\delta L\over \delta A_{36}}&=&\partial_1 \bar {\cal F}^{136}+\partial_2 \bar {\cal F}^{236}+\partial_4 \bar {\cal F}^{436}+\partial_5 \bar {\cal F}^{536}\\
0={\delta L\over \delta A_{45}}&=&\partial_1 \bar {\cal F}^{145}+\partial_2 \bar {\cal F}^{245}+\partial_3 \bar {\cal F}^{345}+\partial_6 \bar {\cal F}^{645}\\
0={\delta L\over \delta A_{46}}&=&\partial_1 \bar {\cal F}^{146}+\partial_2 \bar {\cal F}^{246}+\partial_3 \bar {\cal F}^{346}+\partial_5 \bar {\cal F}^{546}\\
0={\delta L\over \delta A_{56}}&=&\partial_1 \bar {\cal F}^{156}+\partial_2 \bar {\cal F}^{256}+\partial_3 \bar {\cal F}^{356}+\partial_4 \bar {\cal F}^{456}
\ee\\
Above six equations can be expressed as
\be  \partial_a \bar {\cal F}^{abc}=0, ~~~a,b,c\ne 1,2
\ee
which has solution
\be  \bar {\cal F}^{abc}=\epsilon^{12abcd}\partial_d\Phi_{12}
\ee
for arbitrary functions $\Phi^{12}$.  As the gauge symmetry of $\delta A_{12}=\Phi_{12}$ can totally remove $\Phi^{12}$ in ${\cal F}^{abc}$ we have found a self-dual relation
\be {\cal F}_{abc}=0,~~~~~~~a,b,c\ne 1,2
\ee

In the same way, the variation of the action gives 
\be  0={\delta L\over \delta A_{13}}&=&(1-\alpha_1)\partial_4 {\cal F}^{413}+(1-\alpha_2)\partial_5 {\cal F}^{513}+(1-\alpha_3)\partial_6 {\cal F}^{613}\nn\\
&=&\partial_4 \bar{\cal F}^{413}+\partial_5 \bar{\cal F}^{513}+\partial_6 \bar{\cal F}^{613}
\ee
where we have normalized each $\cal F$ by the associated factor $(1-\alpha)$. In the same way we have the relations
\be
0={\delta L\over \delta A_{14}}&=&\partial_3 \bar{\cal F}^{314}+\partial_5 \bar{\cal F}^{514}+\partial_6 \bar{\cal F}^{614}\\
0={\delta L\over \delta A_{15}}&=&\partial_3 \bar{\cal F}^{315}+\partial_4 \bar{\cal F}^{415}+\partial_6 \bar{\cal F}^{615}\\
0={\delta L\over \delta A_{16}}&=&\partial_3 \bar{\cal F}^{316}+\partial_4 \bar{\cal F}^{416}+\partial_5 \bar{\cal F}^{516}
\ee
Above 4 equations give the solution of ${\cal F}^{1ab}$ $(a,b\ne 2)$
\be  {\cal F}^{1ab}=\epsilon^{12abcd}\partial_c\Phi_{d2}
\ee
In the same way, we can find 
\be
 0={\delta L\over \delta A_{23}}&=&\partial_4 \bar{\cal F}^{423}+\partial_5 \bar{\cal F}^{523}+\partial_6 \bar{\cal F}^{623}\\
0={\delta L\over \delta A_{12}}&=&\partial_3 \bar{\cal F}^{324}+\partial_5 \bar{\cal F}^{524}+\partial_6 \bar{\cal F}^{624}\\
0={\delta L\over \delta A_{25}}&=&\partial_3 \bar{\cal F}^{325}+\partial_4 \bar{\cal F}^{425}+\partial_6 \bar{\cal F}^{625}\\
0={\delta L\over \delta A_{26}}&=&\partial_3 \bar{\cal F}^{326}+\partial_4 \bar{\cal F}^{426}+\partial_5 \bar{\cal F}^{526}
\ee
Above 4 equations give the solution of ${\cal F}^{2ab}$ $(a,b\ne 1)$
\be  {\cal F}^{2ab}=\epsilon^{12abcd}\partial_c\Phi_{d1}
\ee
We can now follow [6,7] to find another self-dual relation.  First, taking the Hodge-dual of both sides in above equation we find that 
\be {\cal F}^{1ab}=\partial^{[a}\Phi^{b1]},~~~~(a,b\ne 2)
\ee
Identifying this solution with previou found solution of  ${\cal F}^{1ab}$, then 
\be \partial^{[a}\Phi^{b1]}
=\epsilon^{12abcd}\partial_c\Phi_{d2}
\ee
Acting $\partial_{a}$  on both sides gives
\be \partial_a\partial^a\Phi^{b1}= 0
\ee
under the Lorentz gauge $\partial_a\Phi^{a1}=0$. Now, following [6,7],  imposing the boundary condition that the field $\Phi^{b1}$ be vanished at infinities will lead to the unique solution $\Phi^{b1}=0$ and we arrive at the self-duality conditions
\be {\cal F}^{2ab}=0,~~~~a,b\ne 1
\ee
In the same way, we can find another self-duality conditions
\be {\cal F}^{1ab}=0,~~~~a,b\ne 2
\ee
These complete the proof.
\section {Lorentz Invariance of Self-dual Field Equation}
In this we follow the method of Perry and Schwarz [8] to show that the above general non-covariant actions give field equations with 6d Lorentz invariance.  Note that sec. 4.1, 4.2.1 and 4.2.2 are just those in our previous paper [7], while for completeness we reproduce them in below. 
\subsection {Lorentz transformation of 2-form Field strength}
We first describe the Lorentz transformation of 2-form field strength.  For the coordinate transformation :  $x_a\rightarrow \bar x_a\equiv x_a+\delta x_a$  the tensor  field $H_{\bar M\bar N\bar P}(x_a)$ will becomes 
\be
H_{MNP}(x_a)&\rightarrow &H_{\bar M\bar N\bar P}(x_a+\delta x_a)\equiv{\partial x^Q\over \partial \bar x^{ M}}{\partial x^R\over \partial \bar x^{ N}}{\partial x^S\over \partial \bar x^{ P}}H_{QRS}(x_a+\delta x_a)\nn\\
&\approx&H_{MNP}(x_a+\delta x_a)+{\partial x^Q\over \partial \bar x^{ M}}{\partial x^R\over \partial \bar x^{ N}}{\partial x^S\over \partial \bar x^{ P}}H_{QRS}(x_a)\ee
For the transformation mixing between $x_1$ with $x_\mu$ ($\mu\neq1$) the relation $\delta x_a= \omega_{ab}x^b$ leads to $\delta x_1=-\Lambda_\mu x^\mu$ and  $\delta x_\mu=\Lambda_\mu x^1$ in which we define $\omega_{\mu 1}=-\omega_{1\mu}\equiv \Lambda_\mu$. 

The orbital part of transformation [8] is defined by 
\be \delta_{orb}H_{MNP}&\equiv&H_{MNP}(x_a+\omega_{ab}x^b)-H_{MNP}(x_a)\approx[\delta x_a] \cdot \partial^a H_{MNP}\nn\\
&=&[\Lambda_\mu x^\mu \partial^1] H_{MNP}-x^1[\Lambda_\mu \partial^\mu ] H_{MNP} \nn\\
&=& [(\Lambda\cdot x)\partial^1-  x^1 (\Lambda \cdot\partial)] H_{MNP}\equiv  (\Lambda\cdot L) H_{MNP} 
\ee
Note that $\delta_{orb}$ is independent of index $MNP$ and is universal for any tensor.  

The spin part of transformation [8]  becomes
\be \delta_{spin}H_{\mu\nu\lambda}&\equiv&{\partial x^Q\over \partial x^{\mu}}{\partial x^R\over \partial x^{\nu}}{\partial x^S\over \partial x^{\lambda}}H_{QRS}(x)- H_{\mu\nu\lambda}\nn\\
&=&{\partial (\delta x^1)\over \partial x^{\mu}}{\partial x^R\over \partial x^{\nu}}{\partial  x^S\over \partial x^{\lambda}}H_{1 RS}(x)+{\partial x^Q\over \partial x^{\mu}}{(\delta x^1)\over \partial x^{\nu}}{\partial  x^S\over \partial x^{\lambda}}H_{Q1 S}(x)+{\partial x^Q\over \partial x^{\mu}}{\partial x^R\over \partial x^{\nu}}{\partial  (\delta x^1)\over \partial x^{\lambda}}H_{QR1}(x)
\nn\\
&=&\Big[- \Lambda_\mu H_{1\nu \lambda}\Big]+\Big[-\Lambda_\nu  H_{\mu1\lambda}\Big]+\Big[-\Lambda_\lambda H_{\mu\nu1}\Big]
\ee
and $\delta H_{\mu\nu\lambda}=\delta_{orb}H_{\mu\nu\lambda}+\delta_{spin}H_{\mu\nu\lambda}$

In a same way 
\be \delta_{spin}H_{\mu\nu1}&\equiv&{\partial x^Q\over \partial x^{\mu}}{\partial x^R\over \partial x^{\nu}}{\partial x^S\over \partial x^{1}}H_{QRS}(x)- H_{\mu\nu1}\nn\\
&=&{\partial (\delta x^1)\over \partial x^{\mu}}{\partial x^R\over \partial x^{\nu}}{\partial  x^S\over \partial x^{1}}H_{1 RS}(x)+{\partial x^Q\over \partial x^{\mu}}{(\delta x^1)\over \partial x^{\nu}}{\partial  x^S\over \partial x^{1}}H_{Q1 S}(x)+{\partial x^Q\over \partial x^{\mu}}{\partial x^R\over \partial x^{\nu}}{\partial  (\delta x^\lambda)\over \partial x^{1}}H_{QR\lambda}(x)
\nn\\
&=&0+0+\Lambda^\lambda H_{\mu\nu\lambda}
\ee
and $\delta H_{\mu\nu1}=\delta_{orb}H_{\mu\nu1}+\delta_{spin}H_{\mu\nu1}=(\Lambda\cdot L)H_{\mu\nu1}+\Lambda^\lambda H_{\mu\nu\lambda}$.
\subsection {Lorentz Invariance of  Self-dual Field Equation}
We now use above Lorentz transformation need to examine transformations (I) mixing $x_1$ with $x_2$, (II) mixing $x_1$ with $x_a$ and (IV) mixing $x_a$ with $x_b$, $a,b=3,4,5,6$.
\subsubsection {Mixing $x_1$ with $x_2$}
(I)  Consider first the mixing $x_1$ with $x_2$.  The transformation is
\be \delta x^{1}&=&\omega^{12}~x_2\equiv \Lambda~x_2,~~~\\
\delta x^{2}&=&\omega^{21}~x_{1}=-\Lambda~x_{1}
\ee
Define 
\be \Lambda \cdot L\equiv (\Lambda x_2)\partial_1 -x_1 (\Lambda  \partial_2)
\ee
then
\be
\delta  F_{12 a}&=&(\Lambda \cdot L)F_{12 a}\\
\delta  F_{ a b c}&=&(\Lambda \cdot L)F_{ a b c}\\
\delta  F_{1 a b}&=&(\Lambda \cdot L) F_{1 a b}-\Lambda F_{2ab}\\
\delta  F_{2 a b}&=&(\Lambda \cdot L) F_{2ab}+\Lambda F_{1ab}
\ee 
Using above transformation we can find
\be
\delta \tilde F_{12a}&=&(\Lambda \cdot L) \tilde F_{12 a}+{1\over6}\epsilon_{12 a b c d} (\delta_{spin} F^{ b c d})=(\Lambda \cdot L) \tilde F_{12 a}\\
\delta \tilde F_{1ab}&=&(\Lambda \cdot L) \tilde F_{1 a b}+{1\over6}\epsilon_{1ab2cd}(\delta_{spin} F^{2cd}\cdot 3)\nn\\
&=&(\Lambda \cdot L) \tilde F_{1ab}+{1\over2}\epsilon_{1ab2cd}[\Lambda F^{1cd}]\nn\\
&=&(\Lambda \cdot L) \tilde F_{1ab}-\Lambda\tilde F_{2ab}
\ee 
Thus
\be
\delta (F_{12 a}- \tilde F_{12 a})&=&(\Lambda \cdot L)(F_{12 a}- \tilde F_{12 a})=0\\
\delta (F_{1 a b}-\tilde F_{1 a b})&=&(\Lambda \cdot L) (F_{1 a b}-\tilde F_{1 a b})-\Lambda (F_{2ab}-\tilde F_{2ab})=0
\ee
which are zero for self-dual theory.   Taking Hodge of above relations we also get 
\be
\delta  (F_{abc}- \tilde F_{abc})=0,~~~\delta (F_{2 a b}-\tilde F_{2 a b})=0
\ee
and the non-covariant action gives field equations with 6d Lorentz invariance under transformation mixing $x_1$ with $x_2$. \\
\subsubsection {Mixing $x_1$ with $x_a$}
(II)  For the mixing $x_1$ with $x_{a}$, $a=3,4,5,6$, we shall consider the transformation
\be \delta x^{ a}&=&\omega^{ a 1}~x_1\equiv \Lambda^{ a}~x_1,\\
\delta x^{1}&=&\omega^{1 a }x_{ a}=-\Lambda^{ a}~x_{ a}=-\Lambda \cdot x
\ee
Define 
\be \Lambda \cdot L\equiv (\Lambda \cdot x)\partial_1 -x_1 (\Lambda \cdot \partial)
\ee
then 
\be
\delta  F_{12 a}&=&(\Lambda \cdot L) F_{12 a}+\Lambda^{ b} F_{ b2 a}\\
\delta  F_{ a b c}&=&(\Lambda \cdot L) F_{ a b c}-\Lambda_{ a} F_{1 b c}-\Lambda_{ b} F_{ a 1 c}
-\Lambda_{ c} F_{ a b 1}\\
\delta  F_{1 a b}&=&(\Lambda \cdot L) F_{1 a b}+\Lambda^{ c} F_{ c a b}\\
\delta  F_{2 a b}&=&(\Lambda \cdot L) F_{2 a b}-\Lambda_{ a} F_{21 b }-\Lambda_{ b} F_{2 a1}
\ee 
Use above transformation we can find
\be
\delta  \tilde F_{12 a}&=&(\Lambda \cdot L) \tilde F_{12 a}+{1\over6}\epsilon_{12 a b c d} (\delta_{spin} F^{ b c d})\nn\\
&=&(\Lambda \cdot L) \tilde F_{12 a}+{1\over6}\epsilon_{12 a b c d} [-\Lambda^{ b} F^{1 c d}-\Lambda^{ c} F^{ b1 d}-\Lambda^{ d} F^{ b c1}]\nn\\
&=&(\Lambda \cdot L) \tilde F_{12 a}+\Lambda^{ b} \tilde F_{ a b2}\\
\delta  \tilde F_{1 a b}&=&(\Lambda \cdot L) \tilde F_{1 a b}+{1\over6}\epsilon_{1 a b2 c d}(\delta_{spin} F^{2 c d}\cdot 3)\nn\\
&=&(\Lambda \cdot L) \tilde F_{1 a b}+{1\over2}\epsilon_{1 a b2 c d}[-\Lambda^{ c} F^{21 d}-\Lambda^{ d} F^{ 2c1}]\nn\\
&=&(\Lambda \cdot L) \tilde F_{1 a b}+\Lambda^{ c} \tilde F_{ c a b}
\ee 
Thus 
\be
\delta  (F_{12 a}- \tilde F_{12 a})&=&(\Lambda \cdot L) (F_{12 a}- \tilde F_{12 a})+\Lambda^{ b}(F_{ a b2}- \tilde F_{ a b2})=0\\
\delta (F_{1 a b}-\tilde F_{1 a b})&=&(\Lambda \cdot L) (F_{1 a b}-\tilde F_{1 a b})+\Lambda^{ c} (F_{ c a b}-\tilde F_{ c a b})=0
\ee
which are zero for self-dual theory.  Taking Hodge of above relations we also get 
\be
\delta  (F_{abc}- \tilde F_{abc})=0,~~~\delta (F_{2 a b}-\tilde F_{2 a b})=0
\ee
and the non-covariant action gives field equations with 6d Lorentz invariance under transformation mixing $x_1$ with $x_{ a}$.
\subsubsection {Mixing $x_a$ with $x_b$}
(II)  For the mixing $x_a$ with $x_b$, $a=3,4,5,6$, we shall consider the transformation
\be \delta x^{a}&=&\omega^{ab}~x_b\equiv \Lambda^{ab}~x_b
\ee
Define 
\be \Lambda \cdot L\equiv \Lambda^{ab}(x_a\partial_b -x_b\partial_a)
\ee
then 
\be
\delta  F_{12 a}&=&(\Lambda \cdot L) F_{12 a}-\Lambda_a^{~e} F_{12e}\\
\delta  F_{ a b c}&=&(\Lambda \cdot L) F_{ a b c}-\Lambda_a^{~e}F_{ebc}-\Lambda_b^{~e}F_{aec}-\Lambda_c^{~e}F_{abe}
-\Lambda_{ c} F_{ a b 1}\\
\delta  F_{1 a b}&=&(\Lambda \cdot L) F_{1 a b}-\Lambda_a^{~e}F_{1eb}-\Lambda_b^{~e}F_{1ae}\\
\delta  F_{2ab}&=&(\Lambda \cdot L) F_{2ab}-\Lambda_a^{~e}F_{2eb}-\Lambda_b^{~e}F_{2ae}\ee 
Using above transformations we can find
\be
\delta  \tilde F_{12 a}&=&(\Lambda \cdot L) \tilde F_{12 a}+{1\over6}\epsilon_{12abcd}(\delta_{spin}F^{bcd})\nn\\
&=&(\Lambda \cdot L) \tilde F_{12 a}-{1\over6}\epsilon_{12abcd}(\Lambda^b_eF^{ecd}+\Lambda^c_eF^{bed}+\Lambda^d_eF^{bce})\nn\\
&=&(\Lambda \cdot L) \tilde F_{12 a}-\Lambda_a^{~e} \tilde F_{12e}\\
\delta  \tilde F_{1ab}&=&(\Lambda \cdot L) \tilde F_{1 a b}+{1\over6}\epsilon_{12abcd}(\delta_{spin}F^{2cd}\cdot 3)\nn\\
&=&(\Lambda \cdot L) \tilde F_{1 a b}-{1\over6}\epsilon_{12abcd}(\Lambda^c_eF^{2ed}\cdot 3+\Lambda^d_eF^{2ce}\cdot 3)\nn\\
&=&(\Lambda \cdot L) \tilde F_{1 a b}-\Lambda_a^{~e}\tilde F_{1eb}-\Lambda_b^{~e}\tilde F_{1ae}
\ee 
Therefore 
\be
\delta  (F_{12 a}- \tilde F_{12 a})&=&(\Lambda \cdot L) (F_{12 a}- \tilde F_{12 a})-\Lambda_a^{~e} (F_{12e}-\tilde F_{12e})=0\\
\delta (F_{1 a b}-\tilde F_{1 a b})&=&(\Lambda \cdot L) (F_{1 a b}-\tilde F_{1 a b})-\Lambda_a^{~e}( F_{1eb}-\tilde F_{1eb})-\Lambda_b^{~e}(F_{1ae}-\tilde F_{1ae})=0\nn\\
\ee
which are zero for self-dual theory.  Taking Hodge of above relations we also get 
\be
\delta  (F_{abc}- \tilde F_{abc})=0,~~~\delta (F_{2 a b}-\tilde F_{2 a b})=0
\ee
and the non-covariant action gives field equations with 6d Lorentz invariance under transformation mixing $x_a$ with $x_b$.
\section {10D Self-dual Gauge Theory}
The extension above prescription to 10D self-dual gauge theory is straightforward.  As before, let us first define a  function $L_{ijk\ell mn}$ :
\be L_{ijk\ell m}&\equiv& \tilde F_{ijk\ell m}\times(F^{ijk\ell m}-\tilde F^{ijk\ell m}), ~without~summation~over~indices
\ee
In terms of $L_{ijk\ell m}$ the most general non-covariant  Lagrangian of self-dual gauge theory  is  
\be L_G (\alpha_i)= \sum\limits_a L_{1234a}&+&
({1\over2}+{\alpha_1\over2}) L_{13456}+ ({1\over2}-{\alpha_1\over2}) L_{2789Q}\nn\\
&+&({1\over2}+{\alpha_2\over2}) L_{13457}+ ({1\over2}-{\alpha_2\over2}) L_{26789Q}\nn\\
&+& \cdot\cdot\cdot
\ee
in which $Q$ denotes as tenth dimension hereafter. 
\\\\
{Table 4: Lagrangian in the general decompositions: $D=10$.}\\\\
\scalebox{1}{\hspace{3cm}\includegraphics{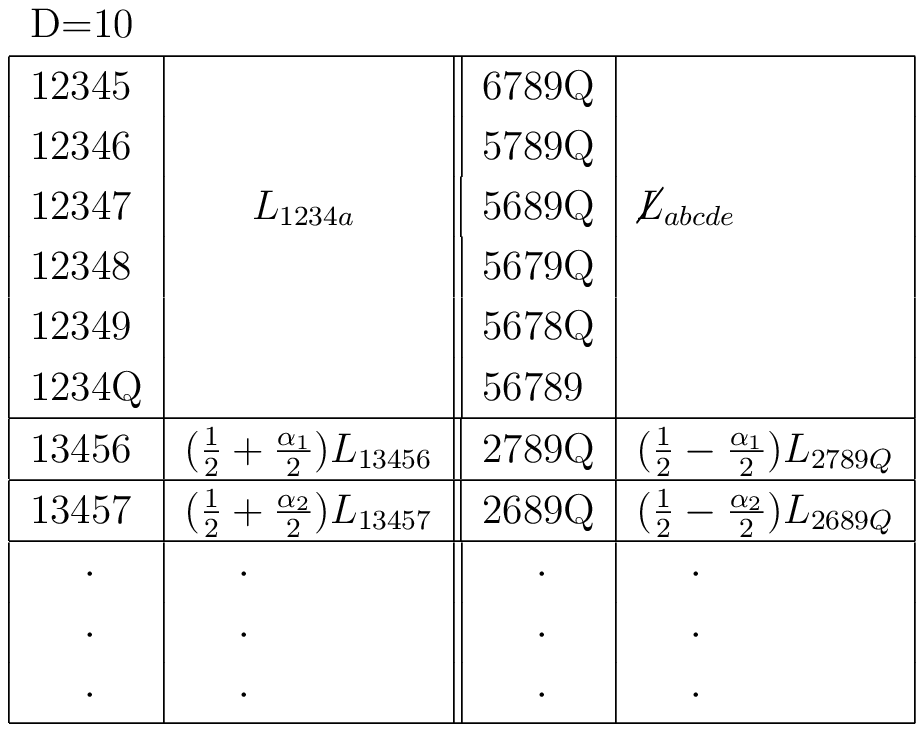}}
\\
\\
From table 4 we see that $L_{G}$ does not picks up $ L_{6789Q},~\cdot\cdot\cdot~,L_{56789}$.  This is a crucial property to have a gauge symmetry.  We now summarize the proof of self-duality of above Lagrangian.
\subsection{Self-duality of 10D Self-dual Gauge Theory}
First,  the variation of the action gives 
\be {\delta L_G (\alpha_i)\over \delta A_{1234}}=\partial_{a}\tilde F^{a1234}=0 
\ee
which is identically zero.  This means that terms involved $A_{1234}$ only through total derivative terms and we have gauge symmetry
\be   \delta A_{1234}&=&\Phi_{1234}
\ee
for arbitrary functions $\Phi_{1234}$.

Next, we can find that 
\be  \partial_a \bar {\cal F}^{abcde}=0, ~~~a,b,c,d,e\ne 1,2,3,4
\ee
which has solution
\be  {\cal F}^{abcde}=\epsilon^{1234abcdef}\partial_f\Phi_{1234}
\ee
for arbitrary functions $\Phi^{1234}$.  As the gauge symmetry of $\delta A_{1234}=\Phi_{1234}$ can totally remove $\Phi^{1234}$ in $ {\cal F}^{abcde}$ we have found a self-dual relation
\be {\cal F}_{abcde}=0,~~~~~~~a,b,c,d,e\ne 1,2,3,4
\ee

In the same way, we can find that
\be  \partial_a \bar {\cal F}^{1abcd}=0, ~~~a,b,c,d\ne 2,3,4
\ee
which has solution
\be  {\cal F}^{1abcd}=\epsilon^{1234abcdef}\partial_f\Phi_{234e} ~~~a,b,c,d\ne 2,3,4
\ee
for arbitrary functions $\Phi^{234e}$.  

We can also find that
\be  \partial_a \bar {\cal F}^{234ab}=0, ~~~a,b\ne 1
\ee
which has solution
\be  {\cal F}^{234ab}=\epsilon^{1234abcdef}\partial_c\Phi_{1def}, ~~~a,b\ne 1
\ee
for arbitrary functions $\Phi^{234e}$.  

We can now follow [6,7] to find another self-dual relation.  First, taking the Hodge-dual of both sides in above equation we find that 
\be {\cal F}^{1abcd}=\partial^{[a}\Phi^{1bcd]},~~~~(a,b,c,d\ne 1,2,3,4)
\ee
Identifying this solution with previoue found solution of  ${\cal F}^{1abcd}$, then 
\be \partial^{[a}\Phi^{1bcd]}
=\epsilon^{1234abcdef}\partial_f\Phi_{2345e},~~~~(a,b,c,d\ne 1,2,3,4)
\ee
Acting $\partial_{a}$  on both sides gives
\be \partial_a\partial^{a}\Phi^{1bcd}= 0,~~~~(a,b,c,d\ne 1,2,3,4)
\ee
under the Lorentz gauge $\partial_a\Phi^{1abc}=0$. Now, following [6,7],  imposing the boundary condition that the field $\Phi^{1bcd}$ be vanished at infinities will lead to the unique solution $\Phi^{1bcd}=0$ and we arrive at the self-duality conditions
\be {\cal F}^{234ab}=0,~~~~a,b\ne 1
\ee
In the same way, we can find all other self-duality conditions. 
\subsection{Lorentz invariance of 10 D Self-dual Field Equation}
 Finally, the method in section IV can be easily applied to prove that general non-covariant actions give field equations with 10d Lorentz invariance.   Essentially, we merely add more index in field strength.

 For example, in considering mixing $x_1$ with $x_2$ we can find that 
\be
\delta  F_{12 abc}&=&(\Lambda \cdot L)F_{12 abc}\\
\delta  F_{ a b cde}&=&(\Lambda \cdot L)F_{ a b cde}\\
\delta  F_{1 a bcd}&=&(\Lambda \cdot L) F_{1 a bcd}-\Lambda F_{2abcd}\\
\delta  F_{2 a bcd}&=&(\Lambda \cdot L) F_{2abcd}+\Lambda F_{1abcd}
\ee 
Using above transformation we can find that 
\be
\delta (F_{12 abc}- \tilde F_{12 abc})&=&(\Lambda \cdot L)(F_{12 abc}- \tilde F_{12 abc})=0\\
\delta (F_{1 a bcd}-\tilde F_{1 a bcd})&=&(\Lambda \cdot L) (F_{1 a bcd}-\tilde F_{1 a bcd})-\Lambda (F_{2abcd}-\tilde F_{2abcd})=0
\ee
which are zero for self-dual theory.   Taking Hodge of above relations we also get 
\be
\delta  (F_{abcde}- \tilde F_{abcde})=0,~~~\delta (F_{2 a bde}-\tilde F_{2 a bde})=0
\ee
and the non-covariant action gives field equations with 10d Lorentz invariance under transformation mixing $x_1$ with $x_2$. 
\section {Conclusion}  
In this paper we have reviewed the various non-covariant formulations  Lagrangian of self-dual gauge theory in 6D and then use the crucial property of the existence of  gauge symmetry $\delta A= \Phi$ to present a general  Lagrangian of  non-covariant forms of self-dual gauge theory.   We have followed previous  prescription [6,7]  to prove the self-dual property in the general Lagrangian.  We furthermore follow the method of Perry and Schwarz [8] to show that the general non-covariant Lagrangian give field equations with 6d Lorentz invariance.    Our method can be straightforward extended to any dimension and we also give a short description about the 10D self-dual gauge theory.  

The result and property found in this paper have shown that there are many kinds of  non-covariant  Lagrangian of self-dual gauge theory and we can easily construct the general Lagrangian.
\\
\\
\begin{center} {\bf REFERENCES}\end{center}
\begin{enumerate}
\item N. Marcus and J.H. Schwarz, Phys. Lett. 115B (1982) 111;\\J. H. Schwarz and A. Sen, ``Duality symmetric actions," Nucl. Phys. B 411, 35 (1994) [arXiv:hep-th/9304154].
\item R. Floreanini and R. Jackiw,``Selfdual fields as charge density solitons,'' Phys. Rev. Lett. 59 (1987) 1873.
\item M. Henneaux and C. Teitelboim, ``Dynamics of chiral (selfdual) p forms,''  Phys. Lett. B 206 (1988) 650.
\item P. M. Ho, Y. Matsuo, ``M5 from M2'', JHEP 0806 (2008) 105 [arXiv: 0804.3629 [hep-th]];  \\P.M. Ho, Y. Imamura, Y. Matsuo, S. Shiba, ``M5-brane in three-form flux and multiple M2-branes,'' JHEP 0808 (2008) 014, [arXiv:0805.2898 [hep-th]].
\item  J. Bagger and N. Lambert, ``Modeling multiple M2'' Phys. Rev. D 75 (2007) 045020 [arXiv:hep-th/0611108];\\
J.A. Bagger and N. Lambert, ``Gauge Symmetry and Supersymmetry of Multiple M2-Branes,'' Phys. Rev. D 77 (2008) 065008 arXiv:0711.0955 [hep-th];\\
J. Bagger and N. Lambert, ``Comments On Multiple M2-branes,'' JHEP 0802 (2008) 105, arXiv:0712.3738 [hep-th]; \\A. Gustavsson, ''Algebraic structures on parallel M2-branes,'' Nucl. Phys. B 811 (2009) 66, arXiv:0709.1260 [hep-th].
\item W.-M. Chen and P.-M. Ho,``Lagrangian Formulations of Self-dual Gauge Theories in Diverse Dimensions'' Nucl. Phys. B837 (2010) 1 [arXiv:1001.3608 [hep-th]]. 
\item Wung-Hong Huang, ``Lagrangian of Self-dual Gauge Fields in Various  Formulations'' Nucl. Phys. BB861 (2012) 403. 
\item M. Perry and J. H. Schwarz, ``Interacting Chiral Gauge Fields in Six Dimensions and Born-Infeld Theory" Nucl.Phys. B489 (1997) 47-64 [arXiv:hep-th/9611065]
\end{enumerate}
\end{document}